\documentclass[useAMS,usenatbib]{mn2e}
\pdfoutput=1
\title[The evolution of the mass ratio of accreting binaries: the role of gas
temperature]{The evolution of the mass ratio of accreting binaries: the role of gas
temperature}
\author[Young et al.]{
M.D.~Young$^1$, 
J.T.~Baird$^1$, 
and
C.J. Clarke$^1$ \\
$^1$Institute of Astronomy, University of Cambridge, Madingley Road, Cambridge,
CB3 0HA, United Kingdom\\
}
\usepackage{graphicx}
\usepackage{natbib}
\usepackage{url}
\usepackage{amsmath}
\usepackage{amssymb}
\usepackage{caption}
\usepackage{subcaption}

\let\baraccent=\= 
\renewcommand{\=}[1]{\stackrel{#1}{=}} 

\begin{document}
\date{Written December 10th 2014}
\maketitle
\begin{abstract}
 We explore an unresolved controversy in the literature about the accuracy of
Smoothed Particle Hydrodynamics (SPH) in modeling the accretion of
gas onto a binary system, a problem with important applications to
the evolution of proto-binaries as well as accreting binary super massive black holes. 
It has previously been suggested that SPH fails to model the flow
of loosely bound material from the secondary to primary Roche lobe and
that its general prediction that accretion drives mass ratios upwards is
numerically flawed. Here we show with 2D SPH that this flow from secondary
to primary Roche lobe is a sensitive function of gas temperature and that
this largely explains the conflicting claims in the literature which
have hitherto been based on either `cold' SPH simulations or `hot'
grid based calculations. We present simulations of a specimen `cold'
and `hot' accretion scenario which are numerically converged and evolved into
a steady state. Our analysis of the conservation of the
Jacobi integral of accreting particles indicates that our results are not
strongly compromised by numerical dissipation.  We also explore the low
resolution limit and find that simulations where the ratio of SPH smoothing
length to disc scale height at the edge of the
circumsecondary is less than $1$ accurately capture binary accretion rates.
\end{abstract}

\section{Introduction}

The accretion of gas onto binary systems is of widespread astrophysical
importance.  For super massive black-hole binaries in galactic
nuclei, the details of gas accretion is important in setting the observational 
signatures of such binaries \citep{BlackHoleSignatures,FlowIntoGap} and in
determining the alignment of black-hole spins \citep{BlackHoleAlign1,BlackHoleAlign2}.

Another important application is to the formation of stellar mass binaries.
This is a central problem for understanding star formation since a large
fraction of stars reside in binary systems
\citep{BfracObs1,BfracObs2,BfracObs3,BfracObs4,BfracObs5,BfracObs6}. 
It is well known that the initial mass of a proto-binary forming from a
collapsing cloud is a small fraction of the mass of the parent cloud
(except for the widest binaries).  The initial proto-binary forms from 
the first material to collapse, which is located near
the clouds centre and thus has low specific angular momentum.  The
proto-binary then acquires most of its mass from the remaining infalling
material, which has greater specific angular momentum.  Because of its greater
specific angular momentum, this material is expected to first form a
circumbinary disc.  The binary then accretes material through inflowing
streams linked to this circumbinary disc.  The ultimate mass
ratio of the binary ($q=M_2/M_1$ where $M_2$ and $M_1$ are 
the masses of the secondary and primary components, respectively) is thus 
set by the relative accretion rates from the circumbinary disc onto the
primary and secondary components, not by the mass ratio of the initial 
proto-binary fragment.

A number of numerical studies have focused on understanding accretion onto
isolated binaries.  These studies have ranged from ballistic calculations
\citep{ArtyBallistic,BateBallistic} to hydrodynamical calculations using
Smoothed Particle Hydrodynamics (SPH) in 3D \citep{BateSPH,Bate_props} and
grid codes in 2D \citep{Ochi,Hanawa,TTauriGrid,ALMA_binary_obs}.  Simulations
of entire clusters have also been used to predict ensemble properties of
binaries \citep{Bate_ClusterSim,Offner_ClusterSim,Goodwin_ClusterSim1,Goodwin_ClusterSim2,Delgado_ClusterSim}.
The resolution of the simulations of entire clusters is inevitably poorer
than in simulations dedicated to a single isolated binary.

SPH (and also ballistic) simulations of circumbinary accretion 
have always found a strong tendency for the majority of the accreted mass to
flow onto the secondary. This is because material inflowing from the
circumbinary disc encounters the secondary first, since it orbits further
from the centre of mass of the binary than the primary. Therefore, SPH studies
of binaries at both the individual binary and cluster scale exhibit a
preference for creating binaries with high $q$ values, since material 
falling preferentially on the secondary implies $\dot{q}>0$.         
This preference is in mild conflict with the observed statistics
of solar mass binaries, which show a flat $q$ distribution \citep{q_dist_disn}.

A re-examination of the accretion of gas onto binaries using 2D grid
based techniques by \cite{Ochi} made the opposite prediction, finding 
more mass accreted by the primary.   
Although these simulations also showed that material encounters the secondary first, 
they found a significant fraction of it continued to flow past the secondary and 
became bound to the primary.  Unless $q << 1$ this results in
a \emph{decreasing} mass ratio (i.e. $\dot{q}<0$).  Higher
resolution grid studies \citep{Hanawa} later confirmed that this effect was not a 
result of insufficient resolution. \cite{Ochi} hypothesised that the discrepancy with 
SPH simulations was due
to excessive numerical dissipation in the SPH code causing accreting particles
to spiral into the secondary instead of transitioning to the primary.

However, as well as using different numerical techniques, the SPH and grid
based studies of accretion onto isolated binaries also differ in the
temperature used for the accreting gas.
Because the dynamical importance of pressure to the flow
of material onto the binary is a function of temperature, it is possible for
gas temperature to qualitatively alter the accretion dynamics.  It
has been suggested that the discrepancy between SPH and grid based
simulations  is due not to numerical technique but to this difference in 
temperature \citep{clarke2008pathways}.

Here we explore the hypothesis that $\dot{q}$ depends on the
temperature of the accreting gas, using 2D SPH simulations at a variety
of resolutions and compare our results with SPH and grid based results
in the literature. We also aim to discover the resolution required
in order to accurately determine the mass flow rates on to the primary and secondary
components of a binary (of value to cluster simulations).  Section \ref{sec:model} 
describes the simulations performed, 
Section \ref{sec:model} reports on the results and 
Section \ref{sec:discussion} discusses the application to
astrophysical systems. Section \ref{sec:conclusion} presents our Conclusions.   

\section{Binary model}
\label{sec:model}

We model a binary on an anti-clockwise circular orbit, 
with mass ratio $q=M_2/M_1$, total mass $M=M_1+M_2$ and separation $a$.  
We use  a modified
version of the SPH code GADGET2 \citep{Gadget2Code} in 2D.  We have chosen 2D
SPH in order to facilitate the comparison with the 2D grid
simulations of \cite{Ochi,Hanawa} and since numerical convergence is faster
in 2D than 3D (for $N$ particles, the SPH smoothing length scales as
$N^{-1/2}$ in 2D and $N^{-1/3}$ in 3D).  
 In a real circumbinary disc, accretion is driven by the effective
viscosity in the disc \citep{DiscReview,MRI,GiantPlanetGravInstab,PP6}
which feeds the binary with material that has similar specific angular momentum 
to that of the binary itself, even if the
material falling into the disc at large radius has a significantly
higher specific angular momentum. In order to avoid the computational
expense of modeling this situation (with numerical or some prescribed
`physical' viscosity driving the secular evolution of the circumbinary
disc) we instead follow previous authors and inject gas  
at large radius $R_{inf} a$ with a ratio ($j_{inf}$) of specific angular momentum
to that of the binary ($\sqrt{GMa}$) of order unity (see Table
\ref{tab:model_params}). We set the initial inward radial velocity of the injected 
particles so that these particles are initially marginally gravitationally 
bound to the binary.
The stars are treated as sink particles so that particles are
accreted if they stray within the accretion radius of each star ($0.01 a$). 
We set the SPH smoothing length $h$ using $h=1.2\sqrt{\frac{m}{\Sigma}}$, where
$m$ is the particle mass and $\Sigma$ is the surface
density \citep{PriceReview}.  This choice gives roughly $18$ neighbours per SPH
particle in 2D.  Details of how to access the exact version of the code used,
along with parameter files and initial conditions can be found in Section
\ref{sec:methods}. 

We employ an isothermal equation of state which is parameterised in terms
of $c$:
\begin{equation}
  c = c_s/\sqrt{\frac{GM}{a}}
  \label{eq:norm_cs}
\end{equation}
which is the dimensionless sound speed (normalised to the orbital speed of
the binary).  We explore binaries with $c=.05$ and $c=.25$. 
For reference, \cite{BateSPH} used $c\le .05$ while \cite{Ochi,Hanawa} used $c\ge 0.18$.  
We neglect both disc self-gravity and the back reaction of the disc on the
binary (i.e. we consider the test particle limit).
The resolution of the simulation is set by parametrising the mass infall rate
in terms of the number of SPH particles
introduced per dynamical timescale of the binary ($\sqrt{{a^3}\over{GM}}$),
so that (at fixed time) the number of particles in the simulation scales
with $\dot{N}$. 

\begin{table}
  \centering
  \begin{tabular}{|l|p{6cm}|}
	 Parameter & Explanation \\
    \hline
    $q=M_2/M_1 = 0.2$ & Mass ratio of the binary. \\
    $j_{\inf} = 1.2$  & Specific angular momentum of injected material
    divided by the specific angular momentum of the binary.\\
    $R_{\inf} = 20 $ & Distance from centre of mass of the binary at which
    material is injected, divided by the binary separation.\\
    $\dot{M}/m = \dot{N}$ & The number of particles injected per
    dynamical time.\\
    $c=\frac{c_s}{\sqrt{\frac{GM}{a}}}$ & Ratio of the isothermal
    sound speed to the orbital speed at one binary separation from the binary
    centre of mass.  Determines the dynamical importance of the pressure
    forces.
  \end{tabular}
  \caption{All the dimensionless parameters which must be specified to
  describe our model.}
  \label{tab:model_params}
\end{table}

Note that the number of injected particles is higher for the `hot' simulations
($c=0.25$: 5-8 in Table \ref{tab:model_params}),
since a significant  fraction of the marginally bound injected
particles are pushed out by pressure gradients. These particles are deleted 
from the simulation soon thereafter.
The `cold'   simulations  (with $c=0.05$: 1-4 in Table \ref{tab:model_params}) 
are each matched with a corresponding hot simulation (5-8) 
so that they have roughly the same number of particles actively contributing
to the simulation at any time (see Table \ref{tab:model_params} for the 
corresponding values of $\dot N$).

\begin{table}
  \centering
  \begin{tabular}{|l|l|l|l|l|}
    ID & $c$ & $\dot{N}$ & $N_{end}$ & $(h/H)_{sec}$ \\
    \hline
    1 & .05 & 100 & 4e4 & .25\\
    2 & .05 & 200 & 8e4 & .1\\
    3 & .05 & 500 & 2e5 & .06\\
    4 & .05 & 1000 & 5e5 & .04\\
    5 & .25 & 200 & 4e4 & .75\\
    6 & .25 & 500 & 1e5 & .45\\
    7 & .25 & 1000 & 2e5 & .3\\
    8 & .25 & 2000 & 5e5 & .2\\
  \end{tabular}
  \caption{Parameters used for simulations in this paper.  $\dot{N}$ is the
  number of particles injected per dynamical time ($\sqrt{a^3/GM}$),
  $N_{end}$ is the number of particles at the end of the simulation and
  $(h/H)_{sec}$ is the resolution at the edge of the disc around the
  secondary.
  All simulations were run with $a=1.0$, $q=0.2$, $j_{\inf}=1.2$,
  $R_{\inf}=20$ and were run for 500 dynamical times.}
  \label{tab:simulations}
\end{table}

For each simulation, we also measure $h/H$ (SPH smoothing length over disc
scale height) at the outer edge of the circumsecondary disc.  This quantity
measures how well resolved the secondary's disc is at the point where it
interacts with infalling material.  As expected, $h/H$ scales as $N^{-1/2}$,
except for the lowest resolution simulations where it increase more
rapidly with declining particle number.  This rapid increase occurs when the 
role of numerical viscosity is enhanced to the point that the accretion time 
scale of the circumsecondary disc becomes comparable with the infall time scale.

\begin{figure}
  \begin{center}
	 \includegraphics[width=0.5\textwidth]{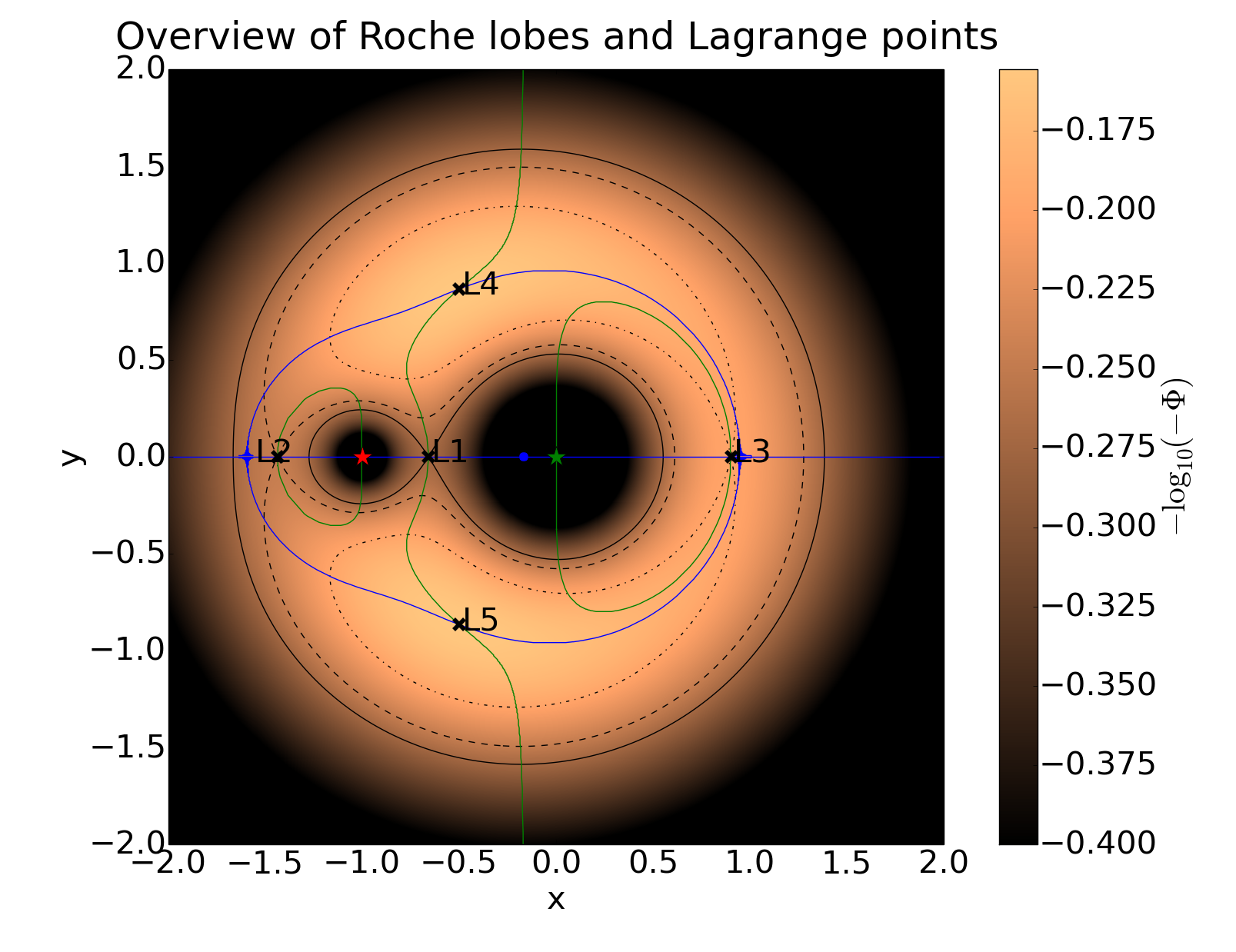}
  \end{center}
  \caption{Equipotential surfaces of $\Phi$, the modified potential
  (gravitational potential plus term to include centrifugal force 
  of non-inertial frame) for the values of $\Phi$ obtained at the three
  Lagrange points L1,L2 and L3.  The colour map shows $\Phi$ elsewhere.  The
  stars are marked in red (secondary) and green (primary) and the centre of
  mass of the system is marked in blue.  The green and blue lines are the
  locations where the x and y components of the force resulting from this
  potential (i.e., $-\nabla{\Phi}$) are zero.}
  \label{fig:Roche_lobes}
\end{figure}

We analyse all our simulations in the frame co-rotating with the binary,
translated so the secondary is at $(x,y) = (-1,0)$ and the primary is at the origin.
In this non-inertial frame it is possible to modify the gravitational potential to
include the effect of the centrifugal force due to the rotating reference
frame.  This modified potential,  the Roche potential, is given by
\begin{equation}
  \Phi =
  -\frac{GM_1}{|{\bf{r}-\bf{r_1}}|}-\frac{GM_2}{|{\bf{r}-\bf{r_2}}|}-\frac{1}{2}(\bf{\Omega}
  \times\bf{r})^2
  \label{eq:mod_pot}
\end{equation}
where $\bf{r}$ is the position vector, $\bf{r_1}$ and $\bf{r_2}$ are the
positions of the primary and secondary respectively and $\bf{\Omega}$ is the
angular momentum vector of the binary.  Equipotential surfaces of $\Phi$, as 
well as points of zero gradient (the Lagrange points) are shown in 
figure \ref{fig:Roche_lobes}.

\section{Results}
\label{sec:res}
\subsection{Dependence of mass ratio evolution on temperature}

\begin{figure*}
  \begin{subfigure}[b]{.49\textwidth}
    \includegraphics[width=\textwidth]{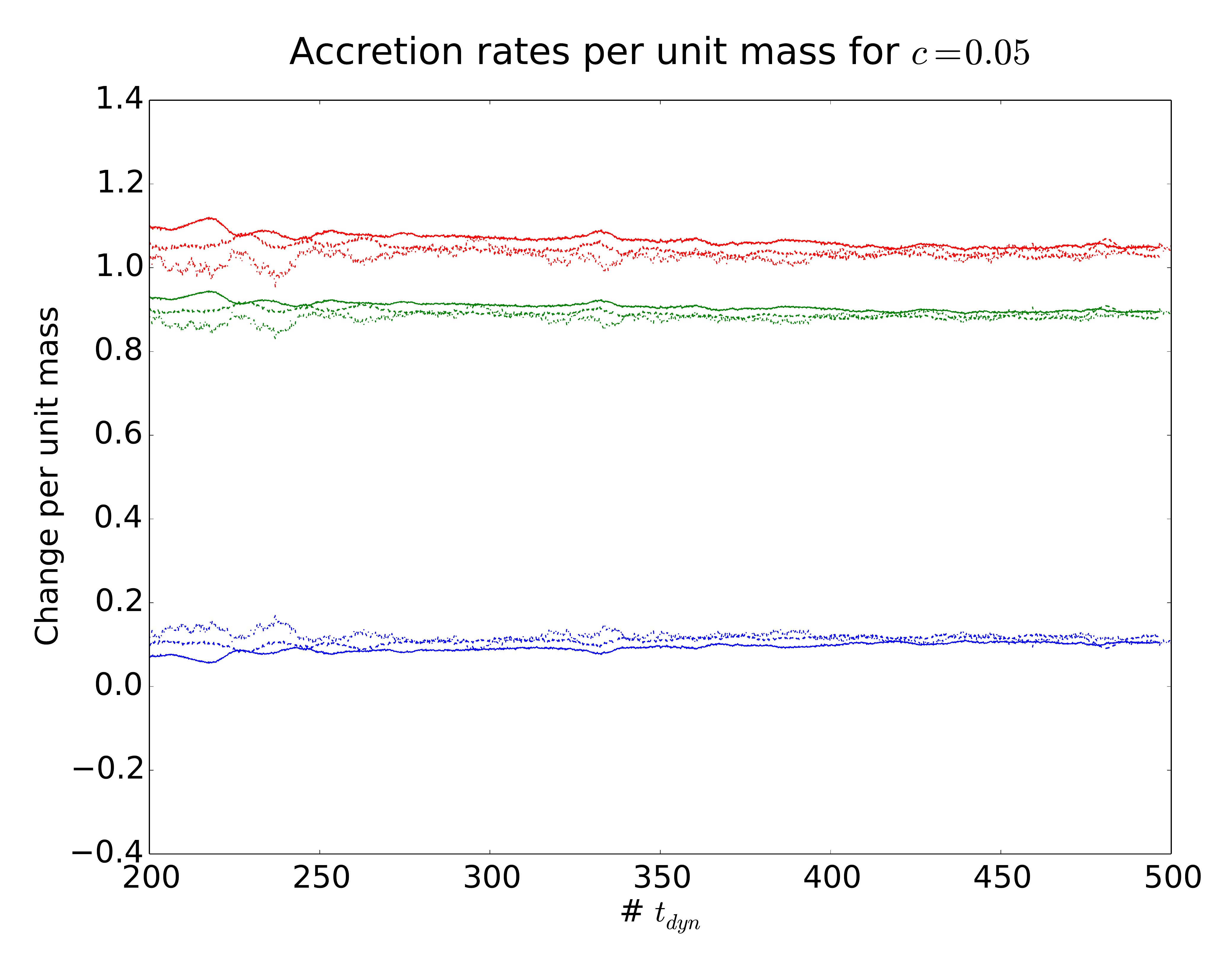}
  \end{subfigure}
  \begin{subfigure}[b]{.49\textwidth}
    \includegraphics[width=\textwidth]{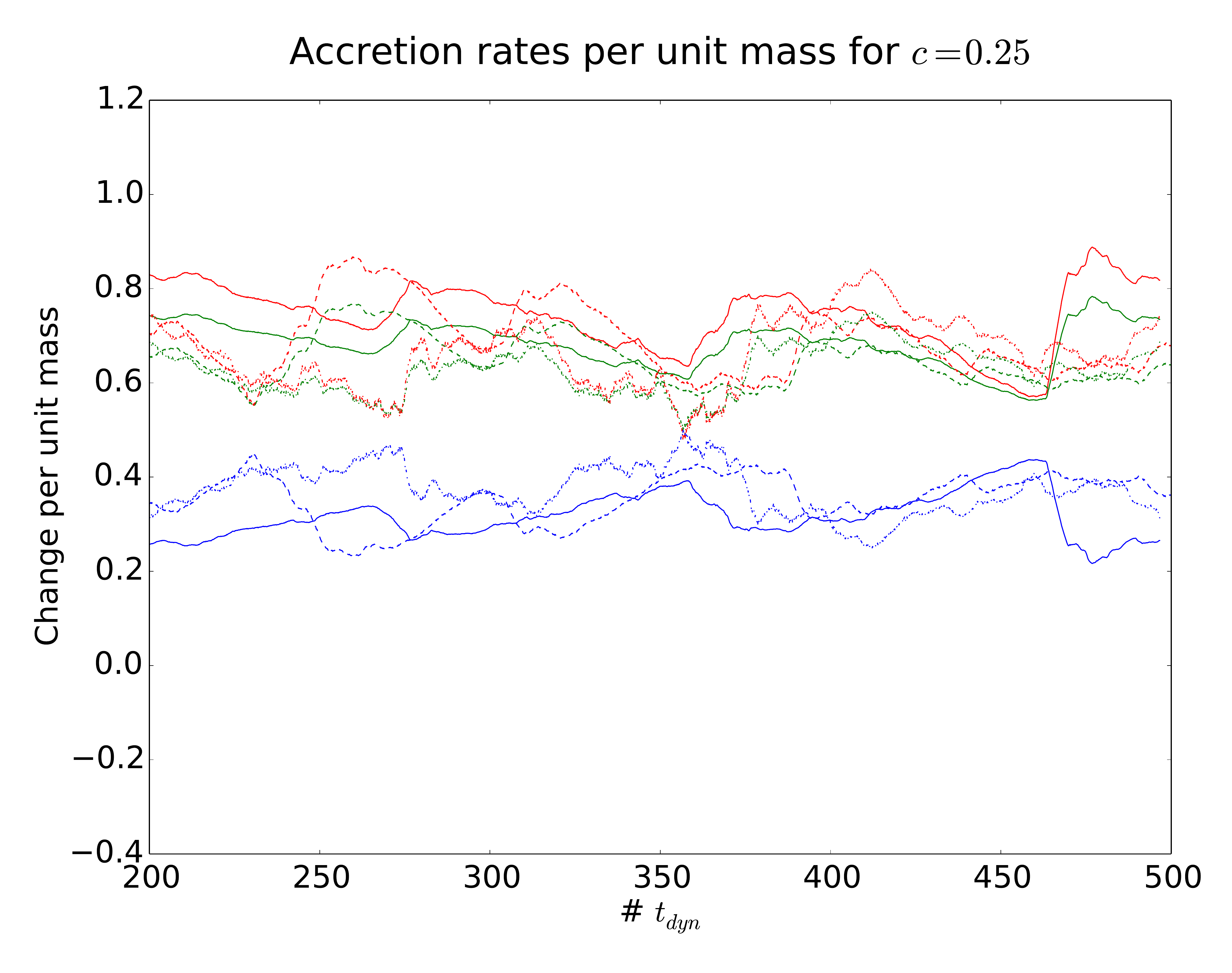}
  \end{subfigure}
  \caption{Accretion rates and corresponding $q\Gamma $ (equation
  \ref{eq:qdot}) for cold (left)
  and hot (right) simulations.  Red lines represent
  $q\Gamma$, blue is accretion onto the primary and green is accretion
  onto the secondary (both per unit accreted mass).  The solid, dashed and
  dot-dashed lines represent $h/H = .04$, $.1$ \& $.25$ ($\dot{N} = 1000,200$ \&
  $100$), respectively, for the
  cold simulation and $h/H = .2$, $.45$ \& .$75$ ($\dot{N}=2000,500$ \& $200$), 
  respectively, for the hot simulation. The first $\sim 30$ orbits are 
  not shown as the simulation has
  yet to settle into a quasi steady-state.}
  \label{fig:conv_accretion_hard}
\end{figure*}

The most important quantity in setting the final, post-accretion, value of $q$
 attained by the binary is the fractional change in $q$ normalised to the
fractional change in mass of the binary.   In our
simulations, this value is given by the dimensionless number:
\begin{equation}
  \Gamma  = \frac{(1+q)}{q(\dot{N_1}+\dot{N_2})}\left( \dot{N_2} - q\dot{N_1} \right)
  \label{eq:qdot}
\end{equation}
where $\dot{N_1}$ and $\dot{N_2}$ are the particle accretion rates of the primary and
secondary (obtained by counting the number of particles accreted by each star). 
To reduce noise, accretion rates were smoothed by taking a rolling average across one 
orbital period of the binary. Figure \ref{fig:conv_accretion_hard}
shows the accretion rates per unit accreted mass, as well as the resulting 
values of $q\Gamma$, for the cold (left panel) and hot (right panel)
simulations.  For both the cold \& hot simulations, simulations are shown at
three resolutions, chosen so they are separated by roughly a factor of $2$ 
in $h/H$ at the edge of the secondary's disc.

It is clear that significantly more material is accreted by the
primary in the hot simulation than in the cold.  Averaging across the steady
state, this translates into a higher value of $\Gamma$ for the cold 
simulation (where $\Gamma \approx 5$) compared with the hot simulation
($\Gamma  \approx 3.5$). We note that for the cold simulation, this
value is
very similar to that found
in the previous 3D SPH simulation of \cite{BateSPH} for the
same parameters, implying that the dimensionality of the problem is
not critical here. 

Our result for the hot simulation is close to, but 
somewhat larger than, the value ($\Gamma \sim 2.5$) obtained by \cite{Ochi} 
for their corresponding model (2-12).  However, the value reported by Ochi
et al is obtained by averaging over a short time period ($t \sim 30$ in the
units of Figure \ref{fig:conv_accretion_hard}) compared to the period of the
variability ($t \sim 90$).  If we average our results on a similarly short
time period, $\Gamma$ can take values between $2.5$ and $4$ depending on when
the average is taken.  Furthermore, the total duration of the Ochi et al simulations 
is shorter than the period we have excluded from our analysis on account of it
not being in a steady state.  As such, the value of $\Gamma$ reported by Ochi 
et al may still
be partially measuring transient accretion driven by their initial conditions.
Finally, Ochi et al use a different definition of accretion (counting mass
within the entire disc) which we found to produce greater variability.  
Given these differences, our results do not suggest a large discrepancy
between grid based and SPH results.

Although we have limited our analysis to 
simulations that have already settled to a quasi steady-state, there is still
some time variability in the accretion rates (particularly in the hot simulations) 
as has been identified by previous authors \citep{FlowIntoGap,Ochi,Hanawa}.
We find very little change in the accretion rates or average $\Gamma$ with
resolution for the models detailed in Table \ref{tab:simulations},
indicating that our results are numerically converged (see Section
\ref{sec:numerics} for an exploration of lower resolution models).  Note also that the
discs become better resolved as a function of time (as more particles
accumulate in our simulation domain) and so the lack of any secular changes
also indicates that numerical convergence has been reached.

\subsection{Dependence of flow morphology on temperature}

\begin{figure*}
  \begin{subfigure}[b]{.99\textwidth}
    \includegraphics[width=\textwidth]{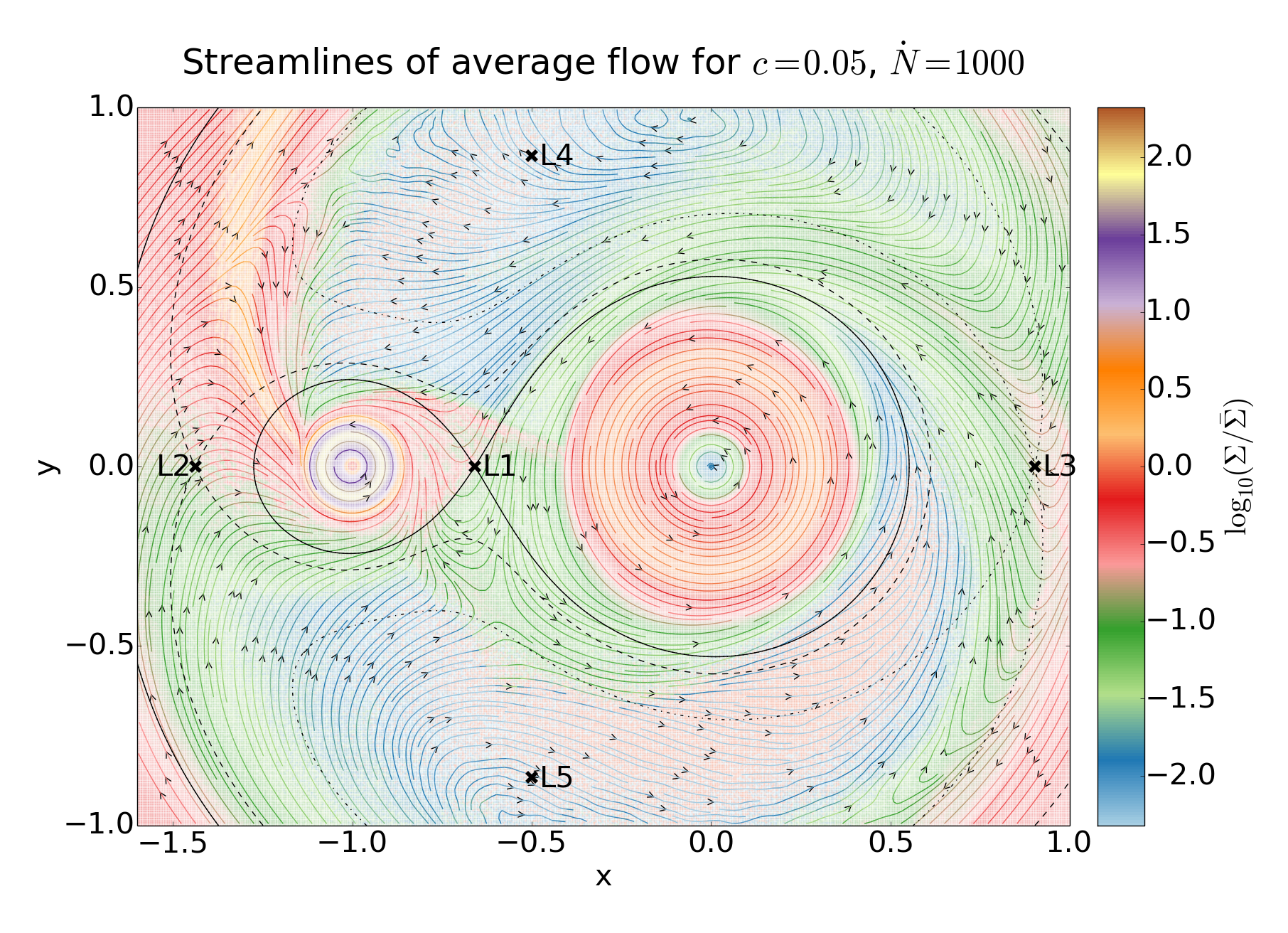}
  \end{subfigure}
  \caption{Average density and streamlines of the velocity field for the
  highest resolution cold simulation (simulation 4).  The density and velocity 
  field were obtained by averaging the velocity and density on a
  grid across roughly 60 orbits.  The colour scheme shows the average 
  surface density, normalised to the surface density in the circumbinary
  disc.  Contours of the Roche potential and the Lagrange points are shown as
  in Figure \ref{fig:Roche_lobes}.}
  \label{fig:streamlines}
\end{figure*}

\begin{figure*}
  \begin{subfigure}[b]{.99\textwidth}
    \includegraphics[width=\textwidth]{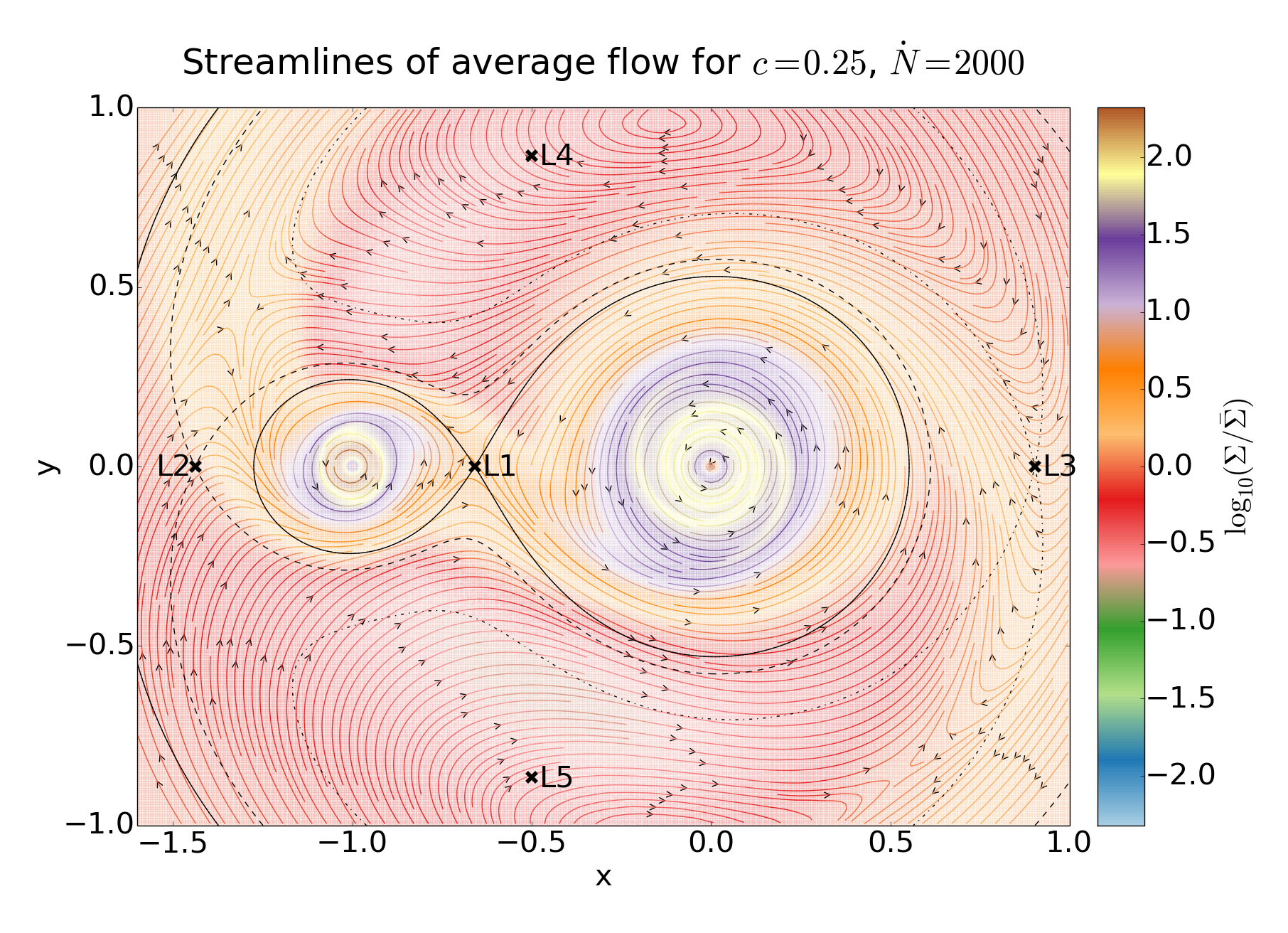}
  \end{subfigure}
  \caption{The same as Figure \ref{fig:streamlines} but for the hot
  simulation (simulation 8).}
  \label{fig:streamlines2}
\end{figure*}

To better understand the flow morphology, we averaged the flow velocity on a
grid over the final 60 binary orbits (the first 30 orbits were ignored as they 
represent a transient settling phase).  Figures \ref{fig:streamlines} and
\ref{fig:streamlines2}
shows streamlines constructed from this averaged flow for the highest
resolution cold and hot simulations respectively.  The colour scheme indicates the average
density of the flow, normalised to the average surface density of the 
circumbinary disc.

In both hot and cold simulations, there are two
spiral arms which join the circumbinary disc to the circumstellar discs
via the L3/L2 Lagrange points. In both cases the spiral arm via L2
is much more pronounced than that through L3. This is consistent with
the lower value of the Roche potential at L2 \citep{Ochi} which
provides an energetically favourable route for material inflowing from
the circumbinary disc. The morphology of the flow in the vicinity of L2
is however quite different in the two cases: in the cold simulation
the flow is narrow and the trajectory of its peak surface density is 
offset from L2 by around $60$ degrees (measured with respect to
the secondary). The flow traverses the Roche lobe (solid line in
Figure \ref{fig:streamlines}) more or less radially and impacts the disc edge well inside
the Roche lobe. At this point the material becomes entrained in the disc
flow. 

In the hot simulation, we find that material of similar
normalised surface density (coded orange) occupies a broad swathe of space between the
secondary and the circumbinary disc and covers nearly half the surface of the
secondary's Roche lobe. 
This much broader structure is simply a result of the much larger pressure scale 
length in the hot simulations.
Moreover, in contrast to the nearly radial flow seen in the cold simulations,
material entering the secondary's Roche Lobe
describes  more or less tangential trajectories, skirting  the boundary of the Roche lobe
`below' the secondary (at negative y).\footnote{This ``skirting'' behaviour is
aided by outwardly directed pressure gradients which produce a force with a
magnitude of $\sim 20\%$ of the inwards pull of gravity.}
At the lower right
of the secondary (at small negative x), the flow divides into a portion
that is retained within the secondary's Roche lobe and one which flows
out through the saddle region in the vicinity of L1. The latter flow of
material then enters the primary's Roche lobe to the lower left of the
 primary. The pronounced orange figure of eight structure which wraps
around the Roche lobes represents weakly bound material which in some
cases orbits the binary several times before eventually being accreted
into the disc of one of the stars. As noted above, the main feeding sources
for this flow are the two spiral arms from the circumbinary disc, especially
that in the region of L2. 

\begin{figure*}
  \begin{subfigure}[b]{.49\textwidth}
    \includegraphics[width=\textwidth]{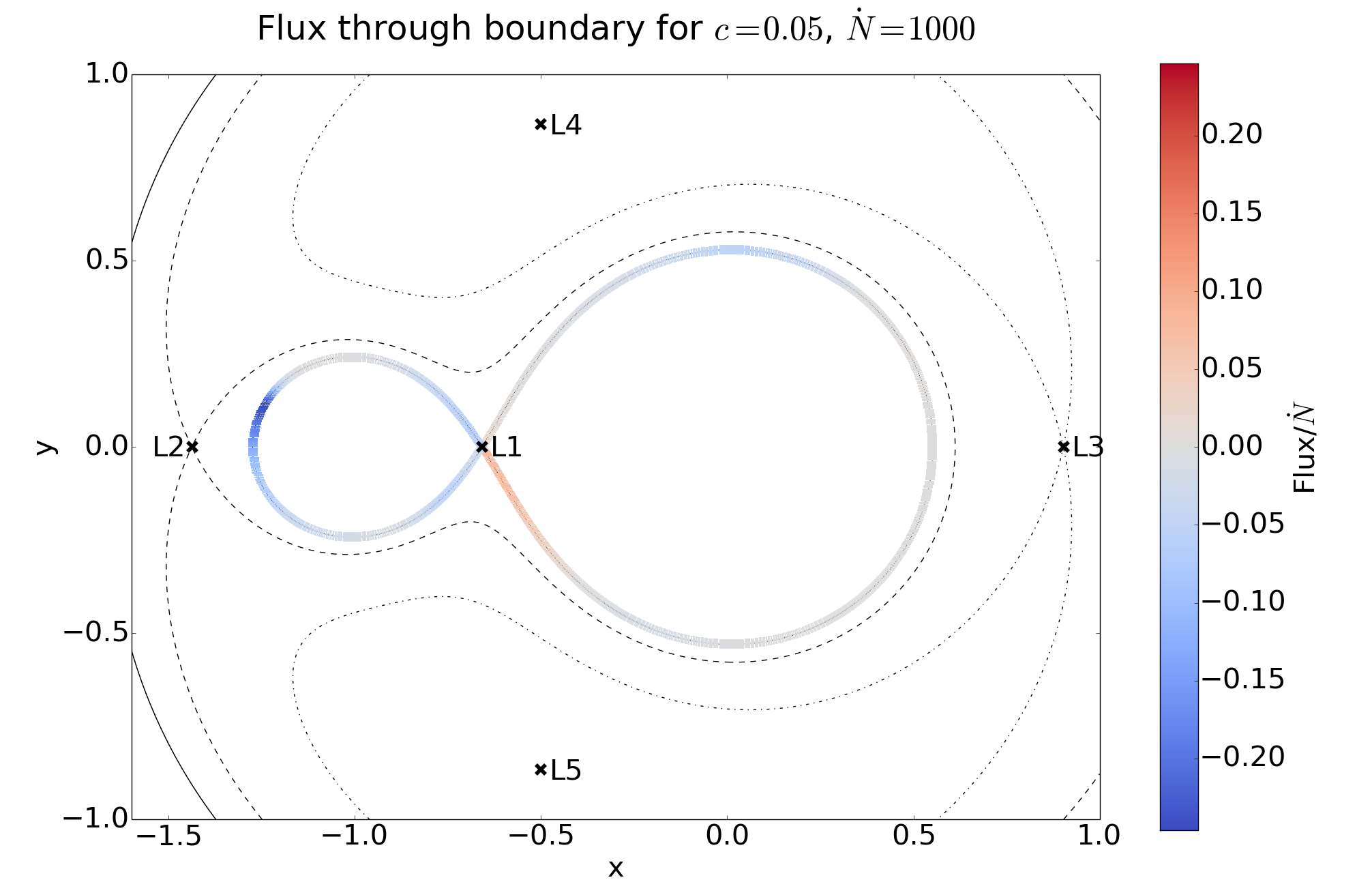}
  \end{subfigure}
  \begin{subfigure}[b]{.49\textwidth}
    \includegraphics[width=\textwidth]{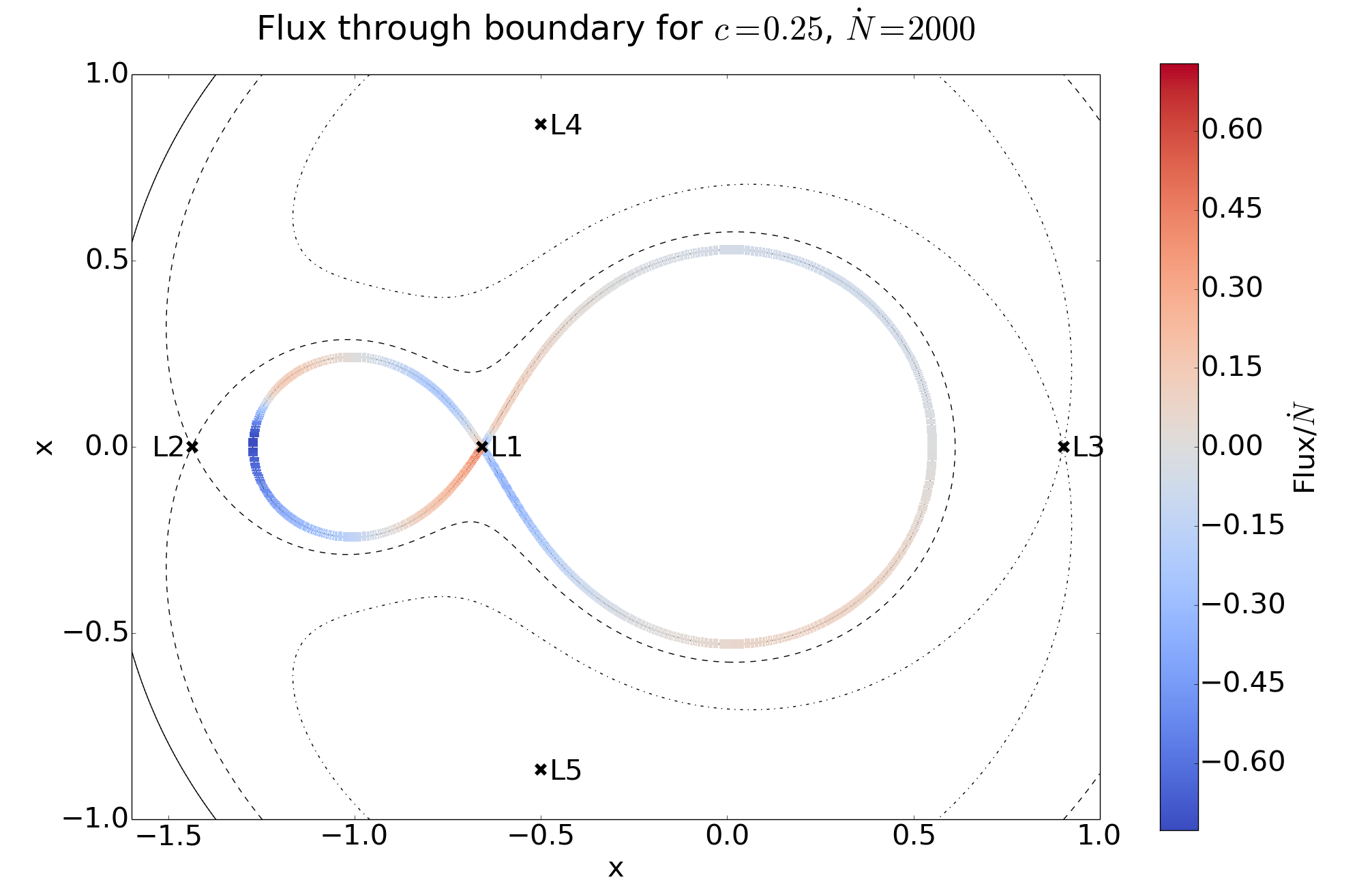}
  \end{subfigure}
  \caption{Flux across Roche lobe boundary for the highest resolution cold
  (left) and hot (right) simulations.  Negative flux implies an inflow.  The
  flux is normalised by the particle injection rate, $\dot{N}$.  See
  text for the details of how the flux is calculated.}
  \label{fig:flux}
\end{figure*}

To further explore the modes of binary accretion demonstrated in
Figures \ref{fig:streamlines} and \ref{fig:streamlines2} we computed the 
mass flux across the boundary of the Roche lobe in our simulations.  In 2D the 
 flux is  $F= \Sigma \bf{v} \bullet d\bf{S}$, which we compute using  SPH
interpolation to estimate $\bf{v}$ \citep{PriceReview,splash} and calculating 
 the  surface normal for the critical Roche equipotential.  This flux was then 
 averaged over the final 60
orbits at each point on the boundary of the Roche lobe and plotted in figure
\ref{fig:flux} for the highest resolution hot and cold simulations. 

Figure \ref{fig:flux} (left) shows that in the
cold simulation the net flux of particles is into the secondary at all points
but is strongly concentrated in the region of the high density feeding
stream (Figure \ref{fig:streamlines}) near L2.  
For the primary, a significant fraction of the inflow  
occurs in the vicinity of (0., 0.5), which can be seen in Figure
\ref{fig:streamlines} to connect
to L3 via a spiral arm. 
The remainder of the primary's Roche lobe is divided into alternating regions of mild 
inflow and mild outflow, of which the most significant is the outflow in the region
of L1.  The outflow near L1 is partly re-accreted by the primary and partly
forms a minor input to the adjacent secondary lobe. 
Over all, the transfer of material between the Roche lobes
is small in the cold simulation, with only $6(18))\%$ of the material
finally accreted by the primary(secondary) having spent some time in
the Roche lobe of the other star.

The picture is quite different in the hot simulations.  
As noted above, the flow into the secondary's Roche lobe in the vicinity
of L2 is much broader and the flow of material once it has entered the Roche
lobe is nearly parallel to the Roche lobe boundary. Although much of this
material is retained
by the secondary, a significant minority exits the secondary's
Roche lobe on the approach to the L1 point and transitions to the
primary's Roche lobe. In the hot simulation this provides the major accretion
stream 
into the primary's Roche lobe (i.e. at around (-0.6,-0.1)). 
Since this flow is nearly parallel to the critical equipotential,
the remainder of the primary's Roche lobe is  again crossed by alternating
bands of mild inflow and outflow; the outflow at small positive
$y$ in the vicinity of L1 is linked to a corresponding mild inflow
to the adjacent portion of the secondary's lobe.

This circulation of material between the lobes does not greatly
affect the dominant accretion mode of the secondary. 
The fraction of material accreting on to the secondary that has ever been in
the primary's Roche lobe is only $21 \%$ in the hot simulation,
a figure scarcely different from that in the cold simulation. However,
 $65 \%$ of the material accreted by
the primary in the hot simulation has spent
time in the secondary's Roche lobe (compared to only $6\%$ in the cold
simulations).

We thus find that the hotter conditions previously modeled by
\cite{Ochi,Hanawa} indeed favour an increased net flow from
secondary to primary.\footnote{We also concur with previous simulations
in finding that the flow morphology and resulting accretion pattern fluctuates
over timescales longer than the orbital timescale due to the shifting location of the
stagnation point near L1.}
This dominant flow
boosts the ratio of accretion  onto the primary to that onto the 
secondary when compared with the cold simulation (see Figure 
\ref{fig:conv_accretion_hard}). 

As noted above, our hot simulation is broadly consistent with
model 2-12 of \cite{Ochi}, a finding which undermines the argument 
that previous `cold' results reported in the literature were
corrupted by numerical issues associated with SPH. Nevertheless we next discuss
whether our results are likely to be corrupted by excessive numerical
dissipation.

\subsection{Numerical effects}
\label{sec:numerics}

The models presented here represent some of the highest resolution
simulations of this problem conducted to date. This statement
may seem surprising given the relatively modest numbers of particles
involved (see Table \ref{tab:simulations}) but is based on an evaluation of the
ratio of a resolution element to the local pressure scale
height \emph{in the vicinity of the Roche lobes} ($h/H$). For our highest
resolution cold simulation the ratio of SPH smoothing
length to pressure scale height in this region is $0.2-0.5$  while
the corresponding value for the previous (3D) cold
SPH simulations of \cite{BateSPH} is of order unity. For our
highest resolution hot simulation this ratio is
$0.04-0.1$, somewhat less than in the grid based simulations
of \cite{Hanawa} and comparable with the simulation of
\cite{Ochi}.\footnote{For grid simulations, we consider the ratio of grid
cell size to pressure scale height.}
We note that grid based codes have not previously been used to
model `cold' conditions, because of the difficulty of resolving the
inner regions of the circumstellar discs in this case.

Given our agreement with lower resolution results in the literature
it is unsurprising that we have demonstrated that our results are
numerically converged.  The SPH smoothing length varies
by around a factor $5$ between models $1/4$ and $5/8$, yet the difference in
fraction accretion rates (shown in Figure \ref{eq:qdot}) is at the $10\%$
level.
This suggests that resolution is \emph{not} a key factor in determining which
component of the binary accretes material. This is difficult to reconcile with
the conjecture of \cite{Ochi} (which ascribed the results from SPH
simulations to excessive numerical dissipation in SPH) since such
dissipation is a function of resolution.
 
As a more direct  test of the effect of numerical dissipation, we follow
\cite{Ochi} and calculate the Jacobi constant $J$ along representative
particle trajectories in our simulations.  For an isothermal equation of
state in 2D, $J$ is given by
\begin{equation}
  J = \frac{1}{2}v^2 + \Phi + c_s^2\ln{\Sigma}
  \label{eq:jacobi}
\end{equation}
where $v$ is the particle speed and $\Phi$ is Roche potential defined in
equation \ref{eq:mod_pot}.  This quantity is just the Bernoulli constant in a
rotating reference frame and so, for an inviscid flow in a steady state, is conserved 
on a per-particle basis.

\begin{figure}
  \begin{center}
	 \includegraphics[width=0.5\textwidth]{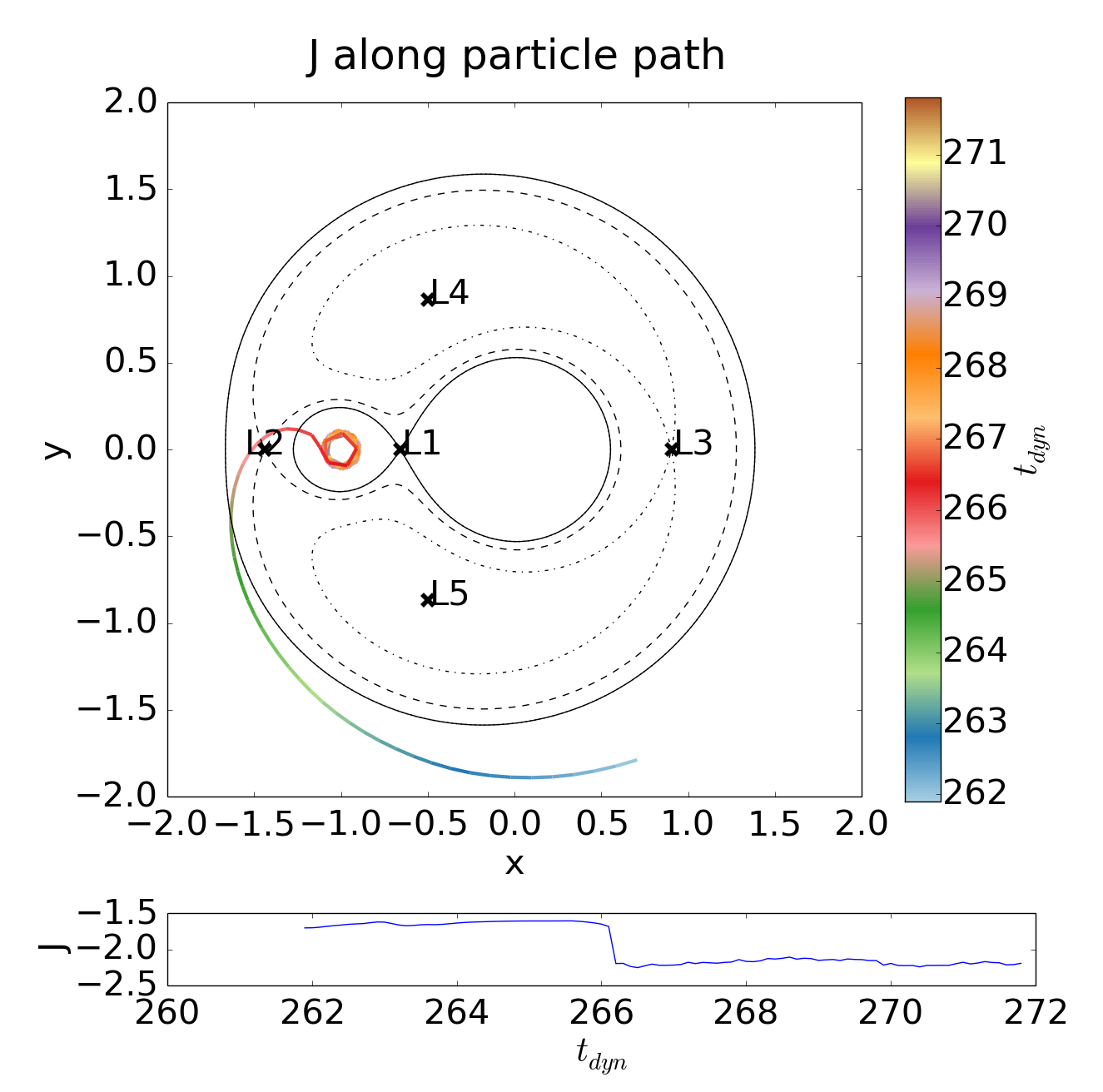}
  \end{center}
  \caption{The Jacobi constant for a single particle at different points along its trajectory (cold simulation).}
  \label{fig:jac_examp}
\end{figure}

We find that for all particles, $J$ only undergoes significant changes on 
timescales shorter than the dynamical time at shock-fronts.  An example is
shown in Figure \ref{fig:jac_examp}, where a particle in the near radial
accretion stream in the cold simulation impacts the secondary's circumstellar
disc.
Away from obvious shock-fronts (such as where the particle in 
Figure \ref{fig:jac_examp} strikes the disc), we find $J$ to be conserved 
to within a few percent on a dynamical
time scale. This means that the material that is skirting the lower edge
of the secondary's Roche lobe in Figure \ref{fig:streamlines2} should change its 
orbital radius due to numerical viscosity by no more than a few per cent in its 
first half-revolution around the secondary. Evidently, such a tiny change would have 
very little effect on which streamlines should remain confined within the 
secondary's Roche lobe and thus on the resulting flow from secondary to
primary. 

We therefore find no evidence that our results are driven by unphysically large
numerical viscosity effects. In order to compare grid based
and SPH results in more detail it will be necessary to compare
numerically converged results for the same parameters and for
simulations that have been run to a quasi-steady state. Such a comparison
is currently not possible since the higher resolution simulations
reported by \cite{Hanawa} are subject to large secular variations
in accretion rates throughout the duration of the experiment. 

\begin{figure}
  \begin{center}
	 \includegraphics[width=0.5\textwidth]{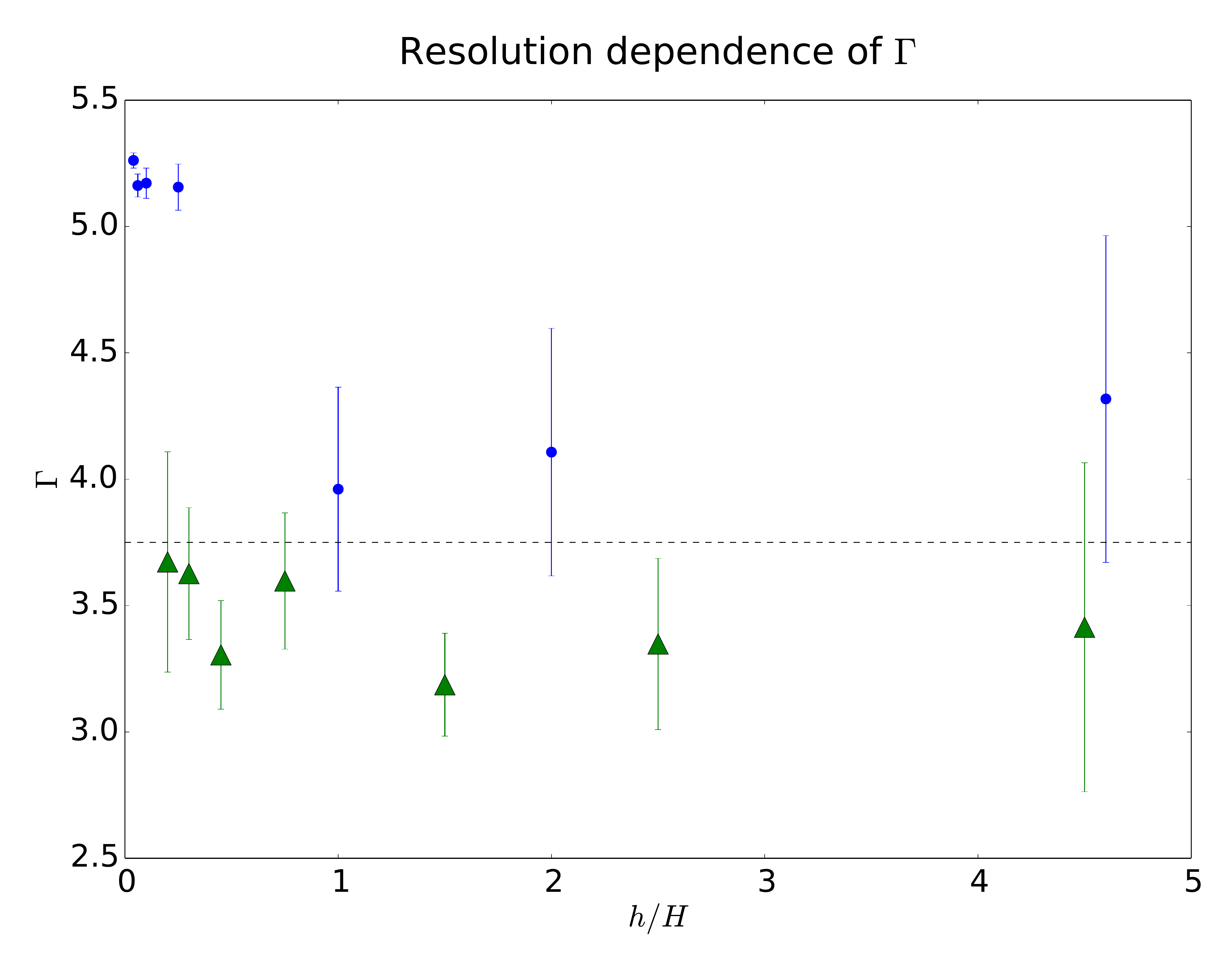}
  \end{center}
  \caption{Average $\Gamma$ for hot (green triangles) and cold (blue circles) simulations,
  averaged over the final 20 binary orbits, for a range of resolutions.  The
  error bars show the variation in $\Gamma$ (1 standard deviation) over the
  same period.
  Resolution is reported as $h/H$ (SPH smoothing length to disc scale height)
  at the edge of the circumsecondary disc.  The dashed line shows the
  $\Gamma$ for purely ballistic particles \citep{BateBallistic}.}
  \label{fig:lowres}
\end{figure}

Finally, we explore the effects of degrading the resolution further, motivated
by a desire to understand the potential systematic effects of under-resolving
binary gas flow in large scale cluster simulations.  In Figure
\ref{fig:lowres} we plot the relative accretion rates onto the primary and
secondary for hot and cold simulations, where we have parameterised the
resolution in terms of the ratio of the SPH smoothing length to pressure scale
length at the edge of the circumsecondary disc.  The lowest resolution
simulations contain only a few hundred particles in the circumsecondary disc.
We include error bars on each simulation to show the range of variation in
$\Gamma$ in the steady-state.

We find that provided $h/H<1$ at the edge of the secondary's disc, the
differential accretion pattern is only modestly affected.  For the cold
simulations, the effect of under-resolving the accretion flow is to
over-estimate accretion onto the primary.  The magnitude of this
over-estimation remains less than $5\%$ so long as $h/H<1$, but once $h/H > 1$
$\Gamma$ rapidly transitions to the value found for the 
accretion of purely ballistic particles \citep{BateBallistic}.
The hot simulations are subject to larger temporal variability in $\Gamma$ 
(see Figure \ref{fig:conv_accretion_hard} and error bars in Figure
\ref{fig:lowres}), but the average $\Gamma$ still only changes by roughly $10\%$
when $h/H<1$.  There is also some evidence for a step change in
$\Gamma$ similar to the cold simulations when $h/H$ exceeds $1$.  
We caution that the fact that the well
resolved hot simulation have $\Gamma$ similar to that of ballistic simulations
makes detecting changes in $\Gamma$ at low resolution potentially difficult
to detect.

We thus conclude that poor resolution in cluster scale simulations (where the
normalised sound speeds tend to be close to our cold case) is unlikely 
to severely compromise the mass ratio of binaries formed in such simulations.

\section{Discussion}
\label{sec:discussion}

Our isothermal simulations  have demonstrated the critical role played by 
the normalised sound speed in determining the relative accretion rates 
on to the primary and secondary. In real (non-isothermal) discs we would 
expect that it is the gas temperature in the vicinity of the Roche lobes that
most critically affects the accretion pattern (since it is here that we have seen
the decisive role of pressure gradients in affecting the
flow morophology). 
This allows us to make some crude estimates of the values of normalised
sound speed to be expected in astrophysical systems.

In the case of proto-binaries, a 1 A.U. binary with primary mass in the regime
typical for T Tauri stars will have orbital velocity $\sim 30$ km s$^{-1}$.
For a typical disc, heated mainly by irradiation from the central star
\citep{TTTemp1,TTTemp2},
we expect $T \sim 100$K for gas on an A.U. scale.  These typical numbers yield 
$c \sim 0.02$.  Since the
temperature of a disc heated by irradiation scales as $R^{-1/2}$, the sound
speed at the relevant distance ($a$) scales as $a^{-1/4}$.  By contrast, the
orbital velocity of the binary scales as $a^{-1/2}$.  Thus the normalised
sound speed scales as $a^{1/4}$.  Eventually (at $\approx 100$ A.U.), the 
temperature attains a `floor' value of order $10$K and the normalised sound
speed rises more steeply with binary separation ($\propto a^{0.5}$).
\footnote{For example, modeling
of the Class I  binary L1551 NE, imaged by ALMA, suggests $c \sim 0.1$
for this wide ($\sim 300$ A.U.) binary.\citep{ALMA_binary_obs}} 

Taken together this suggests
that $c$ lies in the range of $\sim 0.01-0.1$ for proto-binaries 
of roughly solar mass. These values would be a factor of a few larger
or smaller for earlier or later type pre-main sequence stars. 
We thus conclude that $c$ values similar to our cold simulation
are characteristic of young binaries although higher values are possible
in higher mass, wide systems and/or where heating by accretion significantly
boosts the gas temperature.   

 In the case of supermassive black hole binaries, the aspect ratio of
the disc (which is similar to the normalised sound speed in the vicinity
of the binary) is a very weak function of system parameters
and is of order $0.01$ \citep{BHTemp}. Thus the cold regime is the relevant
one in this case also.

\section{Conclusion}
\label{sec:conclusion}

 We have investigated the quantitative and qualitative discrepancies
previously reported in the literature between SPH and grid based 
studies of accretion onto proto-binaries. Our results are broadly
in line with results in the literature obtained by both methods and
demonstrate that the differences reported are in fact due to the different
gas temperatures previously used in SPH and grid based calculations.
The fact that (for a given temperature) our (2D SPH) calculations agree with 
previous 3D SPH calculations and 2D grid based calculations
implies that neither the dimensionality of the simulation nor
the numerical technique are critical in determining the modes of accretion
onto binaries.

 We find that as the ratio of the sound speed in the accreting gas
to the orbital speed of the binary is increased (i.e., the temperature is
increased), there is an increasingly
dominant flow between the secondary and primary Roche lobes. In our
cold and hot simulations (where this ratio is respectively $0.05$ and
$0.25$ ) the fraction of particles ultimately 
accreted by the primary that have previously spent time in the
secondary's Roche lobe changes from $6 \%$ to $65 \%$.  The existence 
of this cross Roche lobe flow accounts
for the slower increase of $q$ in the hotter simulations.

 It would be premature to lay out the implications for protobinary
evolution from this pilot study at a single value of $q$. We have
however demonstrated that it is possible to obtain numerically converged
quasi-steady flow solutions where the role of numerical dissipation
in the critical region around the binary Roche lobes is
demonstrably small away from shocks. We emphasise that 
that
 quantitative statements should not be based on the initial transient
stage of the simulation. We have found that a quasi-steady state
is only attained after $\sim 30$ binary orbits and we base our
accretion rates on long timescale averaging once this initial
phase is concluded.

Finally, we have explored the effect of very poor resolution.  We find that
provided the ratio of the resolution length to the pressure scale height at
the edge of the secondary's disc ($h/H$) remains below 1, the relative accretion 
rates are only mildly compromised.  Our results suggest that even in the poorly
resolved regime expected in cluster scale simulations, the evolution of binary
mass ratios should be captured correctly.

\section{Materials \& Methods}
\label{sec:methods}

In the interests of reproducibility and transparency, all the code needed to
reproduce this work has been made freely available online at
\url{https://bitbucket.org/constantAmateur/binaryaccretion}.  See the readme
file in this repository for further details.

All figures in this paper were generated using the python package
{\sc matplotilb} \citep{MPL}.

\section{Acknowledgements}
\label{sec:ack}

Matthew Young gratefully acknowledges the support of a 
Poynton Cambridge Australia Scholarship.  This work has been 
supported by the DISCSIM project, grant agreement 341137 funded by 
the European Research Council under ERC-2013-ADG. 

We are indebted to Eduardo Delgado-Donate who was working on this problem
at the time of his untimely death in 2007.

We thank the referee for comments that improved the clarity of the paper.

\bibliographystyle{mn2e}
\bibliography{references}

\end{document}